\documentclass[11pt]{article}
\pdfoutput=1 % if your are submitting a pdflatex (i.e. if you have
\usepackage{jheppub} % for details on the use of the package, please
\usepackage[T1]{fontenc} % if needed

\usepackage{euscript,amsmath,amsthm,amsfonts,amssymb}%,show labels}
\usepackage{subcaption}

\title{\boldmath Quantum entanglement and the geometry of spacetime
}

\author{Matthew Headrick}
\affiliation{Martin Fisher School of Physics, Brandeis University, Waltham MA, USA \\
Center for Theoretical Physics, Massachusetts Institute of Technology, Cambridge MA, USA
}

\abstract{
This is a brief, popular-level introduction to holographic entanglement. It was published in the newsletter of the International Centre for Theoretical Sciences, Bangalore.
}

\preprint{BRX-TH-6333, MIT-CTP/5035}
\notoc
\begin{document} 

\maketitle

\newpage
%\flushbottom

Almost since quantum mechanics and general relativity were discovered around a hundred years ago, theoretical physicists have been struggling to unify them. Step by step, this struggle has borne fruit, often in surprising ways and with unexpected implications. Forty years ago, string theory, the first genuine quantum theory of gravity, was found. Twenty years ago, a remarkable discovery came out of string theory: quantum mechanics and general relativity are not actually two separate theories, but rather, in some sense, two sides of the same coin. This connection between quantum mechanics and general relativity goes by the name ``holography''. Ten years ago, theorists studying holography discovered that there is a direct and beautiful relation between entanglement, a central concept in quantum mechanics, and spacetime geometry, a central concept in general relativity. In this article, I will attempt to convey a bit of the story of holographic entanglement, and why it is so exciting.

I will start by explaining a few essential aspects of quantum mechanics and general relativity. Quantum mechanics (QM), which is necessary for an accurate description of systems on atomic and smaller length scales, radically alters our notion of the state of a physical system. While a particle in classical mechanics has, at any instant of time, both a definite position and a definite momentum, according to the famous Heisenberg uncertainty principle in QM it can have neither. This uncertainty leads to the strange phenomenon of entanglement. Entanglement is a kind of correlation between different parts of a quantum system, for example in the positions of different particles. In classical mechanics, correlations between different parts of a system occur only in a statistical sense, when we are ignorant of the true state of the system. Consider, for example, a distribution of possible positions of a pair of particles. If in that distribution the particles are always close to each other, then their positions are statistically correlated. On the other hand, in any actual state, every physical variable has a definite value, including those specifying the states of the parts (for example, the position of each individual particle); since there is no uncertainty, there is no room for correlations. In QM, however, the fact that even in a fixed state physical variables do not take on definite values opens the door to correlations that are intrinsic to the state, rather than reflecting any statistical description or lack of knowledge on our part. Such correlations are called entanglement. Entanglement is at the root of many of the counterintuitive features of QM; this was emphasized by Einstein, who called it ``spooky action at a distance''. It also plays a central role in technologies powered by quantum mechanics, such as quantum cryptography and quantum computation.

We now turn to general relativity (GR). This theory, which is necessary for an accurate description of systems on the scale of the solar system and larger, radically alters our notions of space and time: these are united in a single four-dimensional continuum, whose geometry is variable and dynamic, responding to the matter embedded in it, while dictating how that matter moves. This interaction between spacetime geometry and matter gives rise to the force of gravity.

Although clearly a part of modern physics, GR is labelled by physicists as a ``classical'' theory, by which we mean that it does not obey the rules of QM: the variables specifying the geometry of spacetime, as well as the positions and momenta of particles in spacetime, take on definite values. The problem of combining GR and QM, in other words of finding a consistent quantum theory of gravity, is famously difficult, and has been partly solved by string theory. ``Partly'' means that, while string theory in principle is a fully consistent quantum theory of gravity, and while we can use it to do certain calculations that simultaneously involve QM and GR (for example, involving scattering of gravitons, the particles that carry the force of gravity), we do not yet have a complete understanding of the theory. For this reason, string theory has not yet answered many of the thorniest questions at the intersection of QM and GR, such as the black-hole information puzzle or the nature of the big bang. We also don't yet know whether the real world is actually described by string theory, or by some other quantum theory of gravity.

In 1997, building on investigations of black holes in string theory, Juan Maldacena discovered a very surprising direct connection between QM and GR \cite{Maldacena:1997re}. He showed that certain quantum-mechanical theories are---in a very peculiar sense---also governed by GR. The QM theories in question are similar to the one that governs the strong nuclear force which binds quarks inside atomic nuclei (called quantum chromodynamics). In these theories, spacetime has a fixed geometry and there is no force of gravity. Yet Maldacena showed that they admit a radically different alternative description, which is classical, has a dynamic spacetime geometry, and includes the force of gravity. And there is another important difference: the GR description has an extra dimension of space compared to the QM description. For example, in certain cases the QM system lives in ordinary three-dimensional space while the GR description has four spatial dimensions. In others cases, the QM has two dimensions (lives on a plane), while the GR, like our world, has three. Such a relationship is analogous to a hologram, in which a pattern on a piece of two-dimensional film gives rise to the appearance of a three-dimensional object. For this reason, the correspondence between QM and GR discovered by Maldacena is called ``holographic''.

Before proceeding, we should emphasize one point: It may at first sight seem shocking that a quantum system can ever appear to be classical. In fact, this is the least surprising aspect of holography. After all, any macroscopic object, such as a tennis ball, being composed of electrons and other particles, is strictly speaking governed by QM, yet in describing its motion quantum effects get washed out and classical mechanics is perfectly adequate. The systems considered by Maldacena similarly contain a very large number of physical degrees of freedom; in this case, a very large number of fields. Such systems had already been studied by theorists for decades, and the fact that the collective behavior of these fields is well described by classical mechanics was already anticipated in the 1970s by Gerard 't Hooft \cite{tHooft:1973alw}. However, concrete and useful descriptions of this collective behavior were lacking before Maldacena. The fact that this description in some cases is GR, with an extra dimension of space, was completely unexpected, and remains deeply mysterious.

In the type of quantum theory considered by Maldacena, not only are there a large number of physical degrees of freedom, but they interact very strongly with each other. Theories with strong interactions are very difficult to study by traditional theoretical methods. However, their equivalence to GR makes them relatively tractable, because classical theories are almost always much easier to deal with than quantum ones. Holography has therefore proven incredibly useful for modelling a wide variety of strongly-interacting quantum systems, from colliding atomic nuclei to high-temperature superconductors. Conversely, holography also provides a new perspective on GR and gravity. While the GR theory that governs holographic systems is not exactly the same as the one that governs our universe (in particular, the cosmological constant is negative, in contrast to the observed positive cosmological constant, or ``dark energy''), it is natural to speculate that, even for our universe, the apparently smooth, classical spacetime is really just a representation of a collection of a very large number of strongly interacting quantum-mechanical degrees of freedom; in other words, that space, time, and gravity are emergent phenomena.

\begin{figure}[tbp]
\centering
\includegraphics[width=0.5\textwidth]{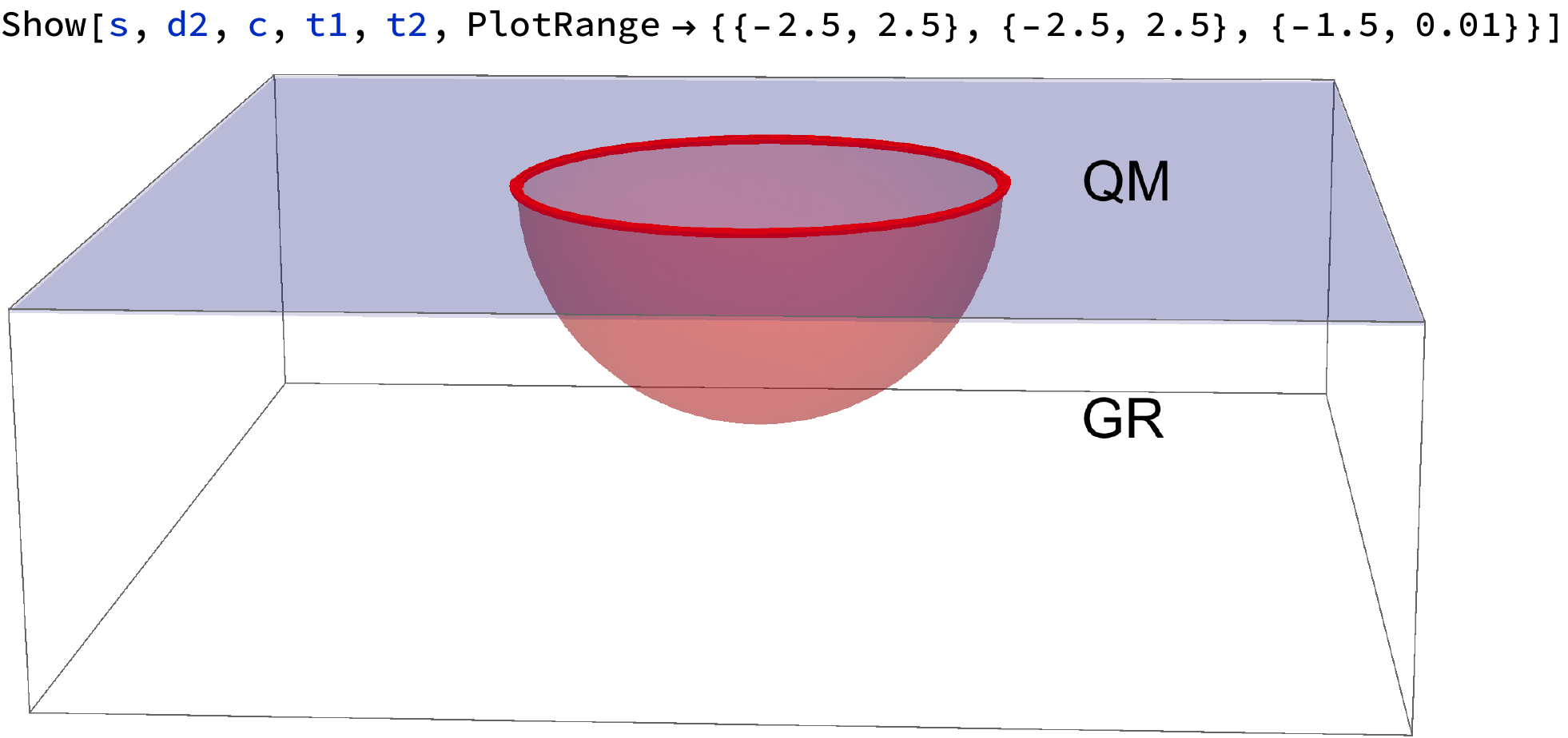}
\caption{%\label{fig:causaldomain}
According to the holographic correspondence, certain two-dimensional quantum-mechanical (QM) systems are equivalent to three-dimensional general relativity (GR). The plane where the QM system lives is the top boundary, shaded in blue, of the GR space. The latter space is warped, with distances being larger than they appear near the top and smaller near the bottom. According to the Ryu-Takayanagi formula, if we divide the QM plane into two parts, then the degree of entanglement between them is given by the area of the minimal surface anchored on their mutual boundary---here, the red circle. Because of the warped geometry, this minimal surface, shown in orange, hangs down rather than stretching flat across the circle as one might expect.
}
\end{figure}

We emphasized above that entanglement is one of the features of QM that most sharply distinguishes it from classical mechanics. If we take a holographic system and consider it in the QM description, then its different parts will naturally be highly entangled with each other. Does that entanglement manifest itself somehow in the GR description? Given that GR is classical and therefore does not admit the possibility of entanglement, one might think that it cannot. Surprisingly, however, the answer is yes. Not only does the entanglement manifest itself in GR, it does so in a beautiful, geometrical way. In 2006, Shinsei Ryu and Tadashi Takayanagi proposed a formula for the degree of entanglement between two parts of a holographic quantum system, quantified by a certain entropy \cite{Ryu:2006bv}. According to their formula, this ``entanglement entropy'' is given by the area of a certain minimal surface in the spatial geometry on the GR side. A minimal surface is one with the smallest possible area subject to some boundary condition; for example, a soap bubble stretched across a loop of wire will arrange itself into a minimal surface. The mathematical problem of finding minimal surfaces is called the Plateaux problem. Ryu and Takayagani thus asserted that, in holographic systems, quantifying entanglement translates into a Plateaux problem. The area here is measured in Planck units, the fundamental units of quantum gravity. In relating the entanglement entropy to a surface area, Ryu and Takayanagi were inspired by the relation discovered by Jacob Bekenstein and Stephen Hawking in the 1970s, giving the entropy of a black hole by the area of its event horizon in Planck units. (See the figure for an illustration of the Ryu-Takayanagi formula.)

Since 2006, the study of holographic entanglement has multiplied in many directions. Since Ryu and Takayanagi essentially guessed their formula, rather than deriving it, a major thrust initially was to check its validity by comparing its predictions both against first-principles calculations and against known general properties of entanglement. It passed all of these tests with flying colors, and this process deepened our understanding of holographic entanglement considerably. Another major thrust has been to generalize the formula, in order to make it as broadly applicable as possible. For example, initially it described only static states, but subsequent work by Veronika Hubeny, Mukund Rangamani, and Takayanagi generalized it to dynamical processes \cite{Hubeny:2007xt}. Their version has been used, for example, to better understand how certain systems thermalize. Both formulas have been applied extensively to holographic models of real-world systems, such as nuclear matter and superconductors, in order to understand their physics better. At a more fundamental level, one school of theorists, led by Mark Van Raamsdonk, has posited that entanglement should be viewed as the basic building block of holography, and attempted to build up GR from this starting point. A notable success has been the derivation of the Einstein equation, the fundamental equation governing GR, from the Ryu-Takayanagi formula \cite{Faulkner:2013ica}.

More generally, Ryu and Takayanagi's discovery has revealed a deep and rich connection between GR and quantum information theory. By now, this connection has been extended far beyond just entanglement to touch on such concepts as quantum error correction \cite{Almheiri:2014lwa}, tensor networks \cite{Swingle:2009bg}, and algorithmic complexity \cite{Brown:2015lvg}. In fact, interesting new developments in quantum information theory have already been spurred by its connections to holography. String theorists and quantum information theorists now routinely meet at conferences and collaborate on papers, a state of affairs that would have been hard to imagine fifteen years ago.

In my view, the discovery of holographic entanglement and its generalizations has been one of the most exciting developments in theoretical physics in this century so far. What other new concepts are waiting to be discovered, and what other unexpected connections? We can't wait to find out.

\acknowledgments{
The author's work was supported by the National Science Foundation through Career Award No.\ PHY-1053842; by the Simons Foundation through \emph{It from Qubit: Simons Collaboration on Quantum Fields, Gravity, and Information} and through a Simons Fellowship in Theoretical Physics; and by the Department of Energy through grant DE-SC0009987. He would also like to thank the MIT Center for Theoretical Physics and the Kavli Institute for Theoretical Physics for hospitality. The KITP is supported in part by the National Science Foundation under Grant No.\ NSF PHY-1748958.
}

\bibliographystyle{JHEP}
\bibliography{refs}

\end{document}